\documentclass[twocolumn,prl,showpacs,showkeys]{revtex4}

\usepackage{amsfonts}
\usepackage{amsmath}
\usepackage{amssymb}
\usepackage{graphicx}
\usepackage{rotating}

\providecommand{\U}[1]{\protect\rule{.1in}{.1in}}
\providecommand{\U}[1]{\protect\rule{.1in}{.1in}}
\providecommand{\U}[1]{\protect\rule{.1in}{.1in}}

\begin{document}

\preprint{}
\title{Anomalous Internal Pair Creation}
\author{P\'{e}ter K\'{a}lm\'{a}n}
\author{Tam\'{a}s Keszthelyi}
\affiliation{Budapest University of Technology and Economics, Institute of Physics,
Budafoki \'{u}t 8. F., H-1521 Budapest, Hungary\ }
\keywords{anomalous internal pair production, nuclear decay}
\pacs{23.20.Ra, 23.90.+w}

\begin{abstract}
In recent electron-positron angular correlation measurements the observed
significant enhancements relative to the internal pair creation at large
angles was interpreted as indication of the creation of $J^{\pi }=1^{+}$
boson called X17 particle. In this paper it is brought up that such
enhancements can be generated by higher order processes. It is found that
nuclear transitions, the transition energy of which is significantly lower
than the whole transition energy, can cause peaked angle dependence in
electron-positron angular correlation.
\end{abstract}

\volumenumber{number}
\issuenumber{number}
\eid{identifier}
\date[Date text]{date}
\received[Received text]{date}
\revised[Revised text]{date}
\accepted[Accepted text]{date}
\published[Published text]{date}
\startpage{1}
\endpage{}
\maketitle

\textit{Introduction.---}The anomalies in the spectra of emitted positrons
observed in heavy-ion collisions at GSI (Darmstadt) in the 1980's \cite%
{Schweppe} - \cite{Cowan2} inspired experimentalists after the suggestions
of \cite{Schafer}, \cite{Balantekin} to search for traces of a short-lived
neutral particle \cite{Savage}, \cite{Savage2} produced in nuclear decays.
The observed significant deviations from internal pair conversion \cite%
{deBoer}, \cite{deBoer2} sustained the interest \cite{deBoer3}, \cite{Vitez}
searching for a light neutral boson \cite{Krasznahorkay3} which is called
X17-boson. Interpreting recent experiments, it was stated that 'to the best
of our knowledge, the observed anomaly can not have a nuclear physics
related origin' \cite{Krasznahorkay1}. In this paper the anomaly is
explained within nuclear physics.

In the experiments \cite{Krasznahorkay1}, \cite{Krasznahorkay2} and \cite%
{Krasznahorkay4} the decay of excited nuclear states through internal
electron ($e^{-}$) - positron ($e^{+}$) pair creation (IPC) was studied. The
examined process was assumed to take place in two successive steps. First,
the excited states of nuclei were prepared in resonant (p,$\gamma $)
reactions. Pair creation is expected after it in a second order
electromagnetic scattering process \cite{Rose1} - \cite{Greiner}. The $%
\Theta $ dependence of the IPC yield fulfilling the $E_{-}+E_{+}=\Delta $
constraint was investigated \cite{Krasznahorkay1}, \cite{Krasznahorkay2}
where $\Theta $ is the angle between the momenta $\mathbf{p}_{-}=\hbar 
\mathbf{k}_{-}$ and $\mathbf{p}_{+}=\hbar \mathbf{k}_{+}$ of the emitted $%
e^{-}$ and $e^{+}$ particles. $E_{-}$, $E_{+}$ and $\mathbf{k}_{-}$, $%
\mathbf{k}_{+}$ are the energies and the wave vectors of $e^{-}$ and $e^{+}$%
, respectively. $\Delta $ is the energy of the resonantly excited transition
and $\hbar $ is the reduced Planck-constant. Extra events, which were said
to be unexplainable with IPC, were found.

It was supposed that paralell with the usual $e^{-}e^{+}$ pair creation,
which is the usual IPC, the decay of the state may also take place by
emitting a hypothetical X17-boson that also decays by $e^{-}e^{+}$ pair
creation having characteristic $\Theta $ dependence. If these extra $%
e^{-}e^{+}$ events originate from the decay of the X17 boson then its rest
mass can be determined with the aid of the given $\Theta $ dependence of the
peaking anomaly appearing around a definite large $\Theta $ angle. Two
different experiments \cite{Krasznahorkay1}, \cite{Krasznahorkay2} resulted
rest masses identical within experimental error with high confidence level 
\cite{Krasznahorkay4}.

However, in the analysis of experiments observing anomalous pair production
the possible effect of higher order processes was not taken into account.
Evaluations are based on the assumption that the populating $p+$ $%
_{Z}^{A}X\rightarrow $ $_{Z+1}^{A+1}Y+\gamma $ capture reaction and the IPC
process take place in two succeeding steps. But higher order coupled
reactions, like the ones to be discussed here and which are one joined
processes contrary to the former two step one, may also happen. In the
higher order processes the creation of the $_{Z+1}^{A+1}Y$ nucleus and the $%
e^{-}e^{+}$ pair are governed by strong and electromagnetic interactions. It
is thought that the anomaly arises if the observed $e^{-}e^{+}$ coincidences
are examined in the light of the two step process only.

\textit{Statement of this Letter.}---The higher order processes, in what
strong and electromagnetic interactions are coupled and govern jointly the
system from the definite initial state to the definite final one, are
investigated. It is shown that they can produce local maximum around a
definite, sometimes large $\Theta $ value in the $\Theta $ dependence of the 
$e^{-}e^{+}$ pair creation yield. Consequently, they may be, at least
partly, responsible for the observed anomalous $e^{-}e^{+}$ pair creation
events.

\textit{General considerations.}---The usual IPC process can be described
with the interaction $U_{EM\text{ }}^{(2)}$, the matrix element $%
\left\langle \nu \right\vert U_{EM\text{ }}^{(2)}$ $\left\vert \mu
\right\rangle \equiv U_{EM,\nu \mu \text{ }}^{(2)}$ of which between states $%
\left\vert \mu \right\rangle $ and $\left\vert \nu \right\rangle $ contains
the Green function $exp\left( iK_{\alpha \beta }R\right) /R$ $\ $where $%
R=\left\vert \mathbf{r}_{e}-\mathbf{r}_{N}\right\vert $ \cite{Rose}, \cite%
{Greiner}. Here $\mathbf{r}_{e}$ and $\mathbf{r}_{N}$ are the
electron/positron and nuclear coordinates and $K_{\alpha \beta }=\left\vert
\Delta E_{\alpha \beta }\right\vert /\left( \hbar c\right) $ is the
transition wavenumber with $\Delta E_{\alpha \beta }$ the change in the
energy of nuclear transition $\alpha \beta $ and $c$ the velocity of light
in vacuum.

Using plane wave $exp(i\mathbf{qr}_{e})$ for the coordinate dependent parts
of the states of the electron-positron pair, and expanding $exp\left(
iK_{\alpha \beta }R\right) /R$ \ and the plane wave in terms of spherical
harmonics \cite{Akhiezer}%
\begin{equation}
U_{EM,\nu \mu \text{ }}^{(2)}\sim \frac{1}{K_{\alpha \beta }\left( K_{\alpha
\beta }^{2}-q^{2}\right) }\left( \frac{q}{K_{\alpha \beta }}\right) ^{L}
\label{UEMnumu2}
\end{equation}%
where $q^{2}=\mathbf{q}^{2}$ with $\mathbf{q}=\mathbf{k}_{+}+\mathbf{k}_{-}$
and $L$ is the multipolarity of the nuclear transition. This characteristics
of $U_{EM,\nu \mu }^{(2)}$, which gives the basis of the $\Theta $
dependence of the yield of $e^{-}e^{+}$ pair creation \cite{Rose1} - \cite%
{Rose}, may help to understand the main features of the $\Theta $ dependence.

In the case of usual IPC, i.e., if $\hbar cK_{\alpha \beta
}=E_{-}+E_{+}=\Delta =\hbar c\delta $, in the denominator of $\left( \ref%
{UEMnumu2}\right) $ the factor $K_{\alpha \beta }^{2}-q^{2}=\delta ^{2}-%
\left[ k_{-}^{2}+k_{+}^{2}+2k_{-}k_{+}\cos \left( \Theta \right) \right] $
increases with increasing $\Theta $ for fixed magnitudes $k_{-}$ and $k_{+}$
of the electron and positron wave vectors. But in higher order processes $%
\left\vert \Delta E_{\alpha \beta }\right\vert \longrightarrow 0$, i.e., $%
K_{\alpha \beta }\longrightarrow 0$ can also happen. In this event $%
K_{\alpha \beta }^{2}-q^{2}\longrightarrow -q^{2}$ and $%
q^{2}=k_{-}^{2}+k_{+}^{2}+2k_{-}k_{+}\cos \left( \Theta \right) $ decreases
with increasing $\Theta $. As a result, the magnitude of the corresponding
matrix-elements decrease in the usual $\left\vert \Delta E_{\alpha \beta
}\right\vert =E_{-}+E_{+}$ case and increase in the $\left\vert \Delta
E_{\alpha \beta }\right\vert \rightarrow 0$ case with increasing $\Theta $.
In the later case, extra $e^{-}e^{+}$ coincidences are expected when $\Theta
\rightarrow 180^{\circ }$.

However, in higher order processes nuclear transitions of $\left\vert \Delta
E_{\alpha \beta }\right\vert $ $<\Delta $ can happen. In these cases the $%
K_{\alpha \beta }^{2}-q^{2}=0$ condition determines the angles%
\begin{equation}
\Theta =\arccos \left[ \frac{K_{\alpha \beta }^{2}-\left(
k_{-}^{2}+k_{+}^{2}\right) }{2k_{-}k_{+}}\right]  \label{thetar}
\end{equation}%
at which singularities appear. For a nuclear transition of given $\hbar
cK_{\alpha \beta }<\Delta $ the minimum angle $\Theta _{m}$ of a possible
singularity arises if $k_{-}=k_{+}$. The angles $\Theta $ of singularities
belonging to the cases $k_{-}\neq k_{+}$ fulfill the condition $\Theta
>\Theta _{m}$. The linked $k_{-}$ and $k_{+}$ values are determined by the $%
E_{-}+E_{+}=\Delta $ condition. The singularities are moderated into peaks
due to the width $\Gamma _{\alpha \beta }=\hbar c\gamma _{\alpha \beta }$ of
the nuclear transition $\alpha \beta $. These peaks may appear in the
transition probability per unit time $W_{\text{fi}}$ of the 3rd or higher
order processes too and they are thought to be responsible for the observed
peaked anomalies in the measured coincident $e^{-}e^{+}$ events. The width $%
\gamma _{\alpha \beta }$ causes the modification of $K_{\alpha \beta }$ as $%
K_{\alpha \beta }\rightarrow $ $K_{\alpha \beta }-i\gamma _{\alpha \beta }/2$
in $\left( \ref{UEMnumu2}\right) $. This modification can also be used in
the results of \cite{Rose1}-\cite{Rose}.

\textit{Transition probability per unit time, }$W_{\text{fi}}$\textit{.---}%
The transition probability per unit time $W_{\text{fi}}$ can be expressed
with the aid of the transition-matrix ($T$-matrix) element $T_{\text{fi}}$ as%
\textit{\ } 
\begin{equation}
W_{\text{fi}}=\frac{2\pi }{\hbar }\sum_{f}\int \int \left\vert T_{\text{fi}%
}\right\vert ^{2}\delta (E)\frac{V^{2}}{\left( 2\pi \right) ^{6}}d\mathbf{k}%
_{+}d\mathbf{k}_{-}  \label{Wfi3}
\end{equation}%
where $\delta (E)=\delta (E_{+}+E_{-}+E_{f}-\Delta )$, $V$ is the volume of
normalization and the sum is made over those final states of energy $E_{f}$
which may contribute to $e^{-}e^{+}$ creation. $T_{\text{fi}}$ may have many
relevant terms, $T_{\text{fi}}^{\text{(3)}}$, $T_{\text{fi}}^{\text{(4)}}$,
etc., which may be responsible for $e^{-}e^{+}$ creation with some of them
for the observed anomalies. $T_{\text{fi}}^{\text{(n)}}$ is the term
obtained in nth order of standard perturbation calculation \cite{Landau4}.
The $T_{\text{fi}}^{\text{(n)}}$ terms can be expressed with the aid of $%
U_{EM\text{ }}^{(2)}$ and $V_{st}$ where $V_{st}$ stands for the potential
of strong interaction. Although in a systematic overview the contribution by
all terms must be taken into account we now focus on the terms which may be
essential in producing extra $e^{-}e^{+}$ events of peaked $\Theta $
dependence.

\textit{Study of the }$T$\textit{-matrix element.}---\ Let us see first a
process, the $T_{\text{fi}}^{\text{(3)}}$ of which can be obtained
adaptating the graphs given in \cite{footnote} changing the interaction of
particles $1$ and $2$ to strong interaction. In this case $T_{\text{fi}}^{%
\text{(3)}}$ itself has many terms. The slowly moving nucleus $_{Z+1}^{A+1}Y$
created by strong interaction and the initial free proton or the free target
nucleus before entering strong interaction may emit $e^{-}e^{+}$ pairs \cite%
{footnote2}. In the corresponding three terms of $T_{\text{fi}}^{\text{(3)}}$
the $K_{\alpha \beta }\longrightarrow 0$ approximation holds leading to $%
exp\left( iK_{\alpha \beta }R\right) /R$ $\longrightarrow 1/R$. These terms
have $1/\left[ k_{-}^{2}+k_{+}^{2}+2k_{-}k_{+}\cos \left( \Theta \right) %
\right] $ like $\Theta $ dependence, which increases with increasing $\Theta 
$. Their effect will not be discussed here.

The strong interaction, which is put in the graphs given by \cite{footnote},
can lead to an excited state $\left\vert n\right\rangle $ of energy $E_{n\nu
}=\varepsilon _{n\nu }-i\Gamma _{n}/2$ where $\Gamma _{n}$ is the width of
the nuclear state $\left\vert n\right\rangle $ of energy distribution $\rho
_{\varepsilon _{n\nu }}=\left[ \Gamma _{n}/\left( 2\pi \right) \right] \left[
\left( \varepsilon _{n\nu }-\varepsilon _{n0}\right) ^{2}+\Gamma _{n}^{2}/4%
\right] ^{-1}$. Here $\varepsilon _{n0}$ is the centre of the distribution.
The energies $\varepsilon _{n\nu }$ and $\varepsilon _{n0}$ are measured
from the energy $E_{f0}$ of the ground state of $_{Z+1}^{A+1}Y$.

In the case of $^{8}Be$ \cite{Krasznahorkay1} two cases of resonant
excitation were studied. These are suffixed with $l=1,2$ further on. The
condition of resonance is determined by rest energies $E_{i0}$ and $E_{f0}$
of the initial and final nuclei, the centre $\varepsilon _{r_{l}0}$ of the
energy distribution of the state, which is tuned to resonance, and the
centre $\epsilon _{0l}$ of the energy of the proton beam as $\varepsilon
_{r_{l}0}=\Delta _{0}+\epsilon _{0l}$ with $\Delta _{0}=E_{i0}-E_{f0}$. Now $%
n=r_{l}$ and $\Delta =\Delta _{0}+\epsilon _{0l}$. Applying the
correspondence $\sum_{\nu }\rightarrow \int \rho _{\varepsilon _{r_{l}\nu
}}d\varepsilon _{r_{l}\nu }$, the relevant $T$-matrix element can be written
as 
\begin{equation}
T_{\text{fi}}^{(3,r_{l})}=U_{EM,fr_{l}\text{ }}^{(2)}V_{st,r_{l}i}\frac{%
\Gamma _{r_{l}}-id_{l}}{\left( d_{l}^{2}+\Gamma _{r_{l}}^{2}\right) }
\label{Tfiar}
\end{equation}%
with $d_{l}$ the detuning and $V_{st,r_{l}i}$ the matrix element of the
strong interaction causing proton capture and resulting resonant transition
into the nuclear state $\left\vert r_{l}\right\rangle $ of $_{Z+1}^{A+1}Y$.
The origin of the detuning $d_{l}\leq D_{l}$ is the energy loss of the
proton beam in the target material of thickness $D_{l}$ usually given in
energy units \cite{Mainsbridge}. The $T_{\text{fi}}^{(3,r_{l})}$ term will
have the dominant $1/\left\{ K_{r_{l}0}^{2}-\left[
k_{-}^{2}+k_{+}^{2}+2k_{-}k_{+}\cos \left( \Theta \right) \right] \right\} $
like behaviour, which decreases with increasing $\Theta $. Its $\Theta $
dependence is identical with the $\Theta $ dependence of the $T$-matrix
element of the second step of the two step process since $K_{r_{l}0}=\Delta
_{0}+\epsilon _{0l}=\Delta $.

In off resonant case%
\begin{equation}
T_{\text{fi}}^{(3,n)}=U_{EM,fn\text{ }}^{(2)}\frac{V_{st,ni}}{i\left(
\varepsilon _{n0}-\Delta _{0}-\epsilon _{0l}\right) },  \label{Tfian}
\end{equation}%
where $V_{st,ni}$ is the matrix element of the strong interaction causing
proton capture and resulting transition into the nuclear state $\left\vert
n\right\rangle $ of $_{Z+1}^{A+1}Y$. The matrix element $T_{\text{fi}%
}^{(3,n)}$ (with $n\neq 0,r_{l}$) of a transition through a non resonant
excited state has peaked $\Theta $ dependence. The peak angle is determined
by $\left( \ref{thetar}\right) $ using $K_{n0}$ in it. (As it was earlier
mentioned, the linked $k_{-}$ and $k_{+}$ values are determined by the $%
E_{-}+E_{+}=\Delta $ condition.)

$e^{-}e^{+}$ pair creation of peaked $\Theta $ dependence can also happen if
nuclear transition takes place between nuclear states $\left\vert
n\right\rangle $ and $\left\vert j\right\rangle $, when the later goes to
the final state due to strong interaction. It is a 4th order process, the $T$%
-matrix element of which reads as%
\begin{equation}
T_{\text{fi}}^{(4,jn)}=V_{st,fj}\frac{U_{EM,jn\text{ }}^{(2)}}{i\varepsilon
_{j0}}\frac{V_{st,ni}}{i\left( \varepsilon _{n0}-\Delta _{0}-\epsilon
_{0l}\right) }.  \label{Tfi4}
\end{equation}%
Since $V_{st}/\varepsilon _{j0}\approx 1$, the magnitude of $T_{\text{fi}%
}^{(4,jn)}$ is comparable with the magnitude of $T_{\text{fi}}^{(3,n)}$.

\textit{Grounds of anomalous }$e^{-}e^{+}$ \textit{creation.---}The
comparison of $\left( \ref{Tfiar}\right) $, $\left( \ref{Tfian}\right) $ and 
$\left( \ref{Tfi4}\right) $ indicates that the leading $T$-matrix element
belongs to the resonant 3rd order process. Its yield can be comparable with
the yield of the two step process since the participation of strong
interaction in a higher order process can compensate for its higher order.
It can be seen from the ratios of values of the astrophysical factors $%
S\left( 0\right) $ of the $d(d,p)^{3}H$, $d(d,\gamma )^{4}He$ reactions
governed by strong interaction and the $^{7}Li(p,\alpha )^{4}He$, $%
^{7}Li(p,\gamma )^{8}Be$ reactions governed by electromagnetic interaction,
which are $10^{3}$ and $40$, respectively \cite{Angulo}. Accordingly, the
contribution to the yield of $e^{-}e^{+}$ coincidences due to the higher
order processes must not be neglected.

The $\Theta _{m}$ values and the transition wavenumbers $K_{n0}$, $K_{nj}$
of a given nucleus are connected via $\left( \ref{thetar}\right) $. In the
case of $^{8}Be$, the preliminary investigation of $\Theta _{m}$ indicates
that besides the actually resonant $T_{\text{fi}}^{(3,r_{l})}=r_{r_{l}}e^{i%
\varphi _{r_{l}}}$ term some $T_{\text{fi}}^{(4,jn)}=r_{jn}e^{i\varphi
_{jn}} $ terms of the $T$-matrix element may be significant. Since in these
cases $\left\vert V_{st,r_{l}i}\left( \Gamma _{r_{l}}-id_{l}\right) /\left(
d_{l}^{2}+\Gamma _{r_{l}}^{2}\right) \right\vert \gg $ $\left\vert
V_{st,ni}/\left( \varepsilon _{n0}-\Delta _{0}-\epsilon _{0l}\right)
\right\vert $ (see $\left( \ref{Tfiar}\right) $, $\left( \ref{Tfi4}\right) $
and $V_{st}/\varepsilon _{j0}\approx 1$) it is also expected that $%
r_{r_{l}}\gg r_{jn}$. These assumptions lead approximately to 
\begin{equation}
\left\vert T_{\text{fi}}^{(3,r_{l})}+\sum_{j,n}T_{\text{fi}%
}^{(4,jn)}\right\vert ^{2}=r_{r_{l}}^{2}+\sum_{j,n}2r_{jn}r_{r_{l}}\cos
(\varphi _{r_{l}}-\varphi _{jn}).  \label{Tfi2}
\end{equation}%
Here, $\varphi _{r_{l}}=\varphi _{r_{l}0}-\arctan (d_{l}/\Gamma _{r_{l}})$.
The $d_{l}$ dependence of $\varphi _{r_{l}}$ indicates that the strength of
the interference term significantly depends on the actual penetration depth
of the proton into the target if the orders of magnitude of $d_{l}$ and $%
\Gamma _{r_{l}}$ are comparable.

In the case of $^{4}He$ the preliminary investigations of $\Theta _{m}$ show
that an other 4th order process, the $r_{jn}$ and $\varphi _{jn}$ values of
which are determined by $\left( \ref{Tfj4}\right) $, may enter into the
approximate expression $\left( \ref{Tfi2}\right) $ of $\left\vert T_{\text{fi%
}}\right\vert ^{2}$. \ In this 4th order process the states $j$ decay to the
ground state with the emission of a soft $E2$ photon of energy $\hbar \omega
\left( E2\right) $ allowed by the energy uncertainty $E_{un}$ of the energy
measurement of the energy sum of the $e^{-}e^{+}$ pair as $\hbar \omega
\left( E2\right) =E_{un}$. The corresponding $T$-matrix element is%
\begin{equation}
T_{\text{fi}}^{(4,jn)}=V_{\gamma ,fj}\frac{U_{EM,j1\text{ }}^{(2)}}{%
i\varepsilon _{j0}}\frac{V_{st,1i}}{\Gamma _{1}},  \label{Tfj4}
\end{equation}%
with $V_{\gamma ,fj}$ the matrix element of $E2$ $\gamma $-coupling and $%
\Gamma _{1}$ the width of state $1$.

The $\Theta $ dependence of extra $e^{-}e^{+}$ pair creation events due to
the term $r_{r_{l}}^{2}$ in $\left( \ref{Tfi2}\right) $ is identical with
the $\Theta $ dependence of the second step of the two step process. The $%
\Theta $ dependence of the remaining terms in $\left( \ref{Tfi2}\right) $ is
of peaked kind. Several transitions of $\hbar cK_{\alpha \beta }<\Delta $
must be taken into account. In consequence of the width $\hbar c\gamma
_{\alpha \beta }$ of the transitions and their appearing range $\Theta
>\Theta _{m}$, the peaks overlap.

\textit{Discussion of anomalies in the} $^{7}Li(p,e^{-}e^{+})^{8}Be$ \textit{%
reaction.---} In the experiment of \cite{Krasznahorkay4} the $%
E_{r_{1}}=17.640$ MeV $\left( 1^{+}\right) $, $\Gamma _{r_{1}}=12.2$ keV and
the $E_{r_{2}}=18.15$ MeV $(1^{+})$, $\Gamma _{r_{2}}=168$ keV states of $%
^{8}Be$ are populated by resonant proton beams of energy $441$ keV ($%
\epsilon _{01}=450$ keV, $D_{1}=9$ keV) and $1030$ keV ($\epsilon _{02}=1100$
keV with $D_{2}=70$ keV) , respectively, with all values in the laboratory
system. The decay of these states through $e^{-}e^{+}$ emission was studied 
\cite{Krasznahorkay1}, \cite{Krasznahorkay4}. The angular $(\Theta )$
distribution of the events fulfilling the $E_{-}+E_{+}$ $=\Delta =\Delta
_{0}+\epsilon _{0l}$ constraint was measured in the case of both resonantly
excited states $\left( l=1,2\right) $. In the case of the $18.15$ MeV state
extra $e^{-}e^{+}$ events peaked at $\Theta \approx 140$ $^{\circ }$ were
observed but in the angular distribution of the events originating from the $%
17.640$ MeV state no peak appeared, although a slight deviation from the
simulated internal pair conversion correlation curve was found at angles
above $110%
{{}^\circ}%
$. The deviation was unstructured and some admix of an $E1$ component
characteristic of the background could explain it. The observation of a peak
at $\Theta \approx 140$ $^{\circ }$ was attributed to the creation and
subsequent $e^{-}e^{+}$ decay of a $J^{\pi }=1^{+}$ boson called X17
particle having rest mass $16.7\pm 0.35$ MeV in the decay of the state of $%
18.15$ MeV energy.

\begin{figure}[tbp]
\resizebox{8.6cm}{!}{\includegraphics*{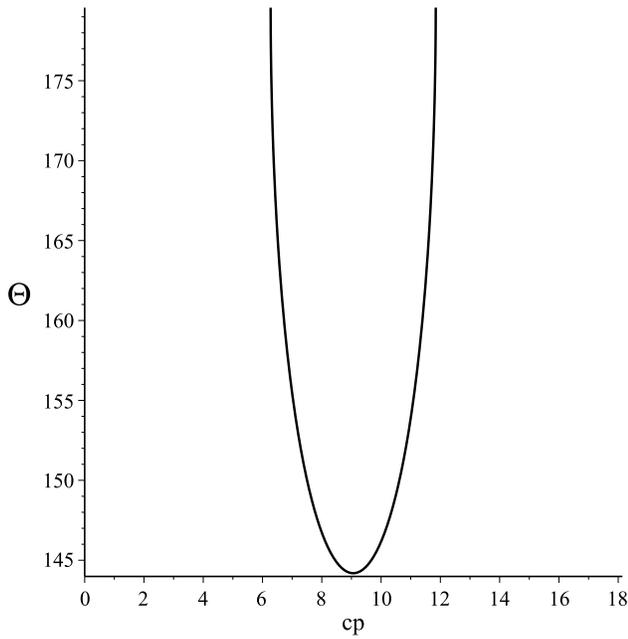}}
\caption{The $cp$ dependence (where $p$ is momentum of the
electron/positron) of $\Theta $ (given by $\left( \protect\ref{thetar}%
\right) $) of the expected peak in the coincident $e^{-}e^{+}$ pair counting
rate in the case of an $E2$ transition of transition energy $\hbar
cK_{31}=5.572$ MeV of $^{8}Be$. $cp$ is measured in MeV units and $\Theta
$ is given in degrees.}
\end{figure}

As it is mentioned above, beside the two step process in both ($17.640$ MeV $%
\left( 1^{+}\right) $, $l=1$ and $18.15$ MeV $(1^{+})$, $l=2$) cases, the $%
T_{\text{fi}}^{(3,r_{l})}$ term of $T_{\text{fi}}^{(3)}$ is dominant. The $%
^{8}Be$ has excited states $E_{1}=11.35$ MeV $\left( 4^{+}\right) $, $%
E_{2}=16.626$ MeV $\left( 2^{+}\right) $, and $E_{3}=16.922$ MeV $\left(
2^{+}\right) $ \cite{Tilley1}. In the processes which are supposed to give
considerable terms to $T_{\text{fi}}^{(4,jn)}$, proton absorption governed
by strong interaction leads to states $2$ or $3$, than $e^{-}e^{+}$ pairs
are created in the $2\rightarrow 1$ or $3\rightarrow 1$ $E2$ transitions.
Finally, strong interaction transition leads to the final state, in which
two $\alpha $ particles of sum energy about $0.09184$ MeV, which is the
decay energy of the ground state of $^{8}Be$ \cite{Shir}, are created. The
values $\hbar cK_{21}=5.276$ MeV and $\hbar cK_{31}=5.572$ MeV if $%
k_{-}=k_{+}$ result from $\left( \ref{thetar}\right) $ that $\Theta
_{2,m}=146.2^{\circ }$ and $\Theta _{3,m}=144.2^{\circ }$, respectively, and
if $k_{-}\neq k_{+}$ that $\Theta _{j,m}<\Theta _{j}<180^{\circ }$ $\left(
j=2,3\right) $ for the angle of the expected peak in the coincident $%
e^{-}e^{+}$ pair counting rate.\ The half of the dominant width $\Gamma
_{1}\approx 3.5$ MeV of state $1$ determines the spread of the peaks as
about $\pm 12^{\circ }$. As an example, the $\Theta _{3}\left( cp\right) $
dependence is plotted in the case of $\hbar cK_{31}=5.572$ MeV in Fig. 1.,
where $p$ is the momentum of either the electron $\left( p_{-}\right) $\ or
positron $\left( p_{+}\right) $. The $E_{-}+E_{+}=\Delta $ constraint
determines the linked $p_{-}=\hbar k_{-}$ and $p_{+}=\hbar k_{+}$ values.
Moreover, the $E_{r_{1}}=17.640$ MeV $\left( 1^{+}\right) $ state can have
an upwards $M1$ coupling to the $E_{4}=27.4941$ MeV $\left( 0^{+}\right) $
state of width $\Gamma =5.5$ keV. The corresponding transition energy is $%
\hbar cK_{40}=9.854$ MeV to which $\Theta _{4,m}=114.1^{\circ }$ belongs. It
may be connected to the observed slight deviation obtained above $110^{\circ
}$ \cite{Krasznahorkay1}.

Supposing that $U_{EM,fr_{1}}^{(2)}V_{st,r_{1}i}$ $\approx
U_{EM,fr_{2}}^{(2)}V_{st,r_{2}i}$ and$\ $emploing $\sqrt{D_{1}^{2}+\Gamma
_{r_{1}}^{2}}\ll $ $\sqrt{D_{2}^{2}+\Gamma _{r_{2}}^{2}}$ in $\left( \ref%
{Tfiar}\right) $, one has $\left\vert T_{\text{fi}}^{(3,r_{1})}\right\vert
\gg \left\vert T_{\text{fi}}^{(3,r_{2})}\right\vert $. Therefore the events
due to the $r_{r_{1}}^{2}$ term (in the case of $l=1$) depress stronger the
events coming from the cross terms $\sum_{n=2,3}2r_{1n}r_{r_{1}}\cos
(\varphi _{r_{1}}-\varphi _{1n})$, which are responsible for the appearance
of peaks, than it does in the case of the state $E_{r_{2}}=18.15$ MeV (in
the case of $l=2$). All the above harmonize well with the observations of 
\cite{Krasznahorkay1}, \cite{Krasznahorkay4}.

\textit{Discussion of anomalies in the} $^{3}H(p,e^{-}e^{+})^{4}He$ \textit{%
reaction.---}In an other work \cite{Krasznahorkay2} the $e^{-}e^{+}$
anomalies in the decay of the $21.01$ MeV $0^{-}$ $\rightarrow 0^{+}$
transition of $^{4}He$ were studied. The second excited state of $^{4}He$ of
energy $21.01$ MeV $\left( 0^{-}\right) $ and center of mass width $\Gamma
=0.84MeV$ \cite{Tilley2}\ was populated in the $^{3}H(p,\gamma )^{4}He$
reaction with a bombarding energy $\epsilon _{p}=900$ keV in the laboratory
frame producing an excitation of $E_{x}=20.49$ MeV of $^{4}He$. In this case
it is stated \cite{Krasznahorkay2} that the measured $e^{-}e^{+}$ angular
correlation anomalies appeared around a peak of a definite angel $115^{\circ
}$. This observation seems to strengthen the X17 boson hypothesis.

The resonant state, the effect of which is taken into account, has energy $%
E_{1}=20.21$ MeV $\left( 0^{+}\right) $ and width $\Gamma _{1}=0.5$ MeV. The 
$^{4}He$ has $2^{+}$ excited states of energy $E_{2}=27.42$ MeV, $%
E_{3}=28.67 $ MeV, $E_{4}=29.89$ MeV and of width $\Gamma _{2}=8.69$ MeV, $%
\Gamma _{3}=3.78$ MeV, $\Gamma _{4}=9.72$ MeV, respectively \cite{Tilley2}.
The $e^{-}e^{+}$ pair is supposed to be created in the $1\rightarrow $ $j$ $%
(j=2,3,4)$ $E2$ transitions. The values $\hbar cK_{21}=7.21$ MeV, $\hbar
cK_{31}=8.46$ MeV and $\hbar cK_{41}=9.68$ MeV result $\Theta
_{2,m}=138.7^{\circ }\pm 26^{\circ }$, $\Theta _{3,m}=131.2^{\circ }\pm
12^{\circ }$ and $\Theta _{4,m}=123.5^{\circ }\pm 32^{\circ }$,
respectively, with $k_{-}=k_{+}$ and $\Theta _{j,m}<\Theta _{j}<180^{\circ }$
$\left( j=2,3,4\right) $ if $k_{-}\neq k_{+}$ for the angle of the expected
peak in the coincident $e^{-}e^{+}$ pair counting rate.\ The spread of $%
\Theta _{j,m}$ is determined by the corresponding $\Gamma _{j}\gg \Gamma
_{1} $ value. As was mentioned above, the energy uncertainty $E_{un}$ of the
energy measurement of the energy sum of the $e^{-}e^{+}$ pair allows to take
into account those processes in which the states $j=2$, $3$ and $4$ decay to
the ground state with the emission of a soft $E2$ photon of energy $\hbar
\omega \left( E2\right) =E_{un}$.

However, similar processes can start from the state of energy $E_{1}=21.01$
MeV $\left( 0^{-}\right) $ and of center of mass width $\Gamma _{1}=0.84MeV$%
. In this case the $1^{-}$ excited states of energy $E_{2}=23.64$ MeV, $%
E_{3}=24.25$ MeV, $E_{4}=25.95$ MeV, $E_{5}=28.37$ MeV and of width $\Gamma
_{2}=6.2$ MeV, $\Gamma _{3}=6.1$ MeV, $\Gamma _{4}=12.66$ MeV, $\Gamma
_{5}=3.92$ MeV, respectively, \cite{Tilley2} are coupled to state $1$ with $%
M1$ coupling and the states $j=2,...,5$ decay emitting a soft $M1$ photon of
energy $\hbar \omega \left( M1\right) =E_{un}$. But the process can also
take place through these intermediate states starting from the $E_{1}=20.21$
MeV $\left( 0^{+}\right) $ state with $E1$ coupling to them and by emission
of a final soft $E1$ photon from these states. Moreover, the $E_{1}=21.01$
MeV $\left( 0^{-}\right) $ may have $E1$ coupling with the state of energy $%
E_{5}=28.31$ MeV $\left( 1^{+}\right) $ and of width $\Gamma _{2}=6.2$ MeV
too. All the corresponding $\Theta _{j,m}$ values can be determined as well.
Thus in this case a great number of reactions can lead to $e^{-}e^{+}$
anomalies.

\textit{Summary.---} It was raised that $e^{-}e^{+}$ anomalies to the usual
IPC decay of an excited nuclear state can be ascribed to reactions of higher
order of standard perturbation calculation. The observed anomalous peak \cite%
{Krasznahorkay1}, \cite{Krasznahorkay4} is well explained in the case of
decay of resonantly excited state of $^{8}Be$. Qualitative explanation of
recent anomalous $e^{-}e^{+}$ observations \cite{Krasznahorkay2}, \cite%
{Krasznahorkay4} made in the case of the decay of resonantly excited states
of $^{4}He$ is also presented. It is concluded that the assumption of X17
particle does not seem to be necessary to explain the observed $e^{-}e^{+}$
anomalies.


\begin{thebibliography}{99}
\bibitem{Schweppe} J. Schweppe \textit{et al.}, Phys. Rev. Lett. \textbf{51}%
, 2261-2264 (1983).

\bibitem{Clemente} M. Clemente, E. Berdermann, P. Kienle, H. Tsertos, W.
Wagner, C. Kozhuharov, F. Bosch, and W. Koenig, Phys. Lett. B \textbf{137},
41-46 (1984).

\bibitem{Cowan} T. Cowan \textit{et al.}, Phys. Rev. Lett. \textbf{54},
1761-1764 (1985).

\bibitem{Cowan2} T. Cowan \textit{et al.}, Phys. Rev. Lett. \textbf{56},
444-447 (1986).

\bibitem{Schafer} A. Sch\"{a}fer, J. Reinhardt, B. M\"{u}ller, W. Greiner,
and G. Soff, J. Phys. G : Nucl. Phys. \textbf{11}, L69-L74 (1985).

\bibitem{Balantekin} A. B. Balantekin, C. Bottcher, M. R. Strayer, and S. J.
Lee, Phys. Rev. Lett. \textbf{55}, 461-464 (1985).

\bibitem{Savage} M. J. Savage, R. D. McKeown, B. W. Filippone and L. W.
Mitchell, Phys. Rev. Lett. \textbf{57}, 178-181 (1986).

\bibitem{Savage2} M. J. Savage, B. W. Filippone and L. W. Mitchell, Phys.
Rev. D. \textbf{37}, 1134-1141 (1988).

\bibitem{deBoer} F. W. N. de Boer \textit{et al.}, Phys. Lett. B \textbf{388}%
, 235-240 (1996).

\bibitem{deBoer2} F. W. N. de Boer, R. van Dantzigz, J. van Klinken, K.
Bethge, H. Bokemeyer, A. Buda, K. A. M\"{u}ller, and K. E. Stiebing, J.
Phys. G: Nucl. Part. Phys. \textbf{23}, L85--L96 (1997).

\bibitem{deBoer3} F. W. N. de Boer, K. Bethge, H. Bokemeyer, R. van Dantzig,
J. van Klinken, V. Mironov, K. A. M\"{u}ller, and K. E. Stiebing, J. Phys.
G: Nucl. Part. Phys. \textbf{27}, L29--L40 (2001).

\bibitem{Vitez} A. Vit\'{e}z, A. Krasznahorkay, J. Guly\'{a}s, M. Csatl\'{o}%
s, L. Csige, Z. G\'{a}csi, A. Krasznahorkay Jr., and B. M. Nyak\'{o}, Acta
Physica Polonica B \textbf{39}, 483-487 (2008).

\bibitem{Krasznahorkay3} A. Krasznahorkay \textit{et al.}, Frascati Physics
Series \textbf{56}, 86-97 (2012).

\bibitem{Krasznahorkay1} A. J. Krasznahorkay \textit{et al.}, Phys. Rev.
Lett. \textbf{116}, 042501 (2016).

\bibitem{Krasznahorkay2} A. J. Krasznahorkay \textit{et al.}, arXiv:
1910.10459.

\bibitem{Krasznahorkay4} A. J. Krasznahorkay \textit{et al.}, Acta Physica
Polonica B \textbf{50}, 675-684 (2019).

\bibitem{Rose1} M. E. Rose, Phys. Rev. \textbf{76}, 678-681 (1949).

\bibitem{Goldring} G. Goldring, Proc. Phys. Soc. A \textbf{66}, 341- 345
(1953).

\bibitem{Rose} M. E. Rose, Phys. Rev. \textbf{131}, 1260-1264 (1963).

\bibitem{Greiner} P. Schl\"{u}ter, G. Soff, and W. Greiner, Phys. Rep. 
\textbf{75}, 327-392 (1981).

\bibitem{Akhiezer} A. I. Akhiezer and V. B. Berestetskii, \textit{Quantum
Electrodynamics }(Interscience Publishers-Wiley, New York, 1965).

\bibitem{Landau4} V. B. Berestetskii, E. M. Lifschitz, and L. P. Pitaevskii, 
\textit{Quantum Electrodynamics, }2nd edn. in \textit{Course of Theoretical
Physics,} Vol. 4. (Pergamon Press, Oxford-New York, 1982).

\bibitem{footnote} See Fig. (100.6), p. 446 of \cite{Landau4}.

\bibitem{footnote2} The terminology 'before' is used corresponding to time
ordering of perturbation calculation of quantum mechanics.

\bibitem{Mainsbridge} B. Mainsbridge, Nucl. Phys. \textbf{21}, 1-14 (1960).

\bibitem{Angulo} C. Angulo\textit{\ et al.}, Nucl.Phys. A \textbf{656},
3-183 (1999).

\bibitem{Tilley1} D. R. Tilley, J. H. Kelley, J. L. Godwin, D. J. Millener,
J. E. Purcell, C. G. Sheu, and H. R. Weller, Nucl. Phys. A \textbf{745},
155--362 (2004).

\bibitem{Shir} R. B. Firestone and V. S. Shirly, \textit{Tables of Isotopes}%
, 8th ed. (Wiley, New York, 1996).

\bibitem{Tilley2} D. R. Tilley, H. R. Weller, and G. M. Hale, Nucl. Phys. A 
\textbf{541}, 1--104 (1992).
\end{thebibliography}
\end{document}